\documentclass[titlepage,floatfix,floats,superscriptaddress,reprint,jcp,aip,showpacs]{revtex4-1}
\pdfoutput=1
\usepackage{amssymb}
\usepackage{amsmath}
\usepackage{amsfonts}
\usepackage{graphicx}
\usepackage[mathscr]{euscript}
\usepackage{natbib}
\usepackage{color}

\definecolor{orange}{rgb}{1,0.5,0}

\begin{document}
\title{A variational conformational dynamics  approach to the selection of collective variables in metadynamics} 
\author{James McCarty}
\affiliation{Department of Chemistry and Applied Biosciences, ETH Z{\"u}rich c/o USI Campus,
             6900 Lugano, Switzerland.}
\affiliation{Facolt\'{a} di Informatica, Instituto di Scienze Computazionali, and National Center for Computational Design and Discovery of Novel Materials MARVEL, Universit\'{a} della Svizzera italiana, 6900 Lugano, Switzerland}
\author{Michele Parrinello}
\email{parrinello@phys.chem.ethz.ch}
\affiliation{Department of Chemistry and Applied Biosciences, ETH Z{\"u}rich c/o USI Campus,
             6900 Lugano, Switzerland.}
\affiliation{Facolt\'{a} di Informatica, Instituto di Scienze Computazionali, and National Center for Computational Design and Discovery of Novel Materials MARVEL, Universit\'{a} della Svizzera italiana, 6900 Lugano, Switzerland}

\date{\today}
\newpage
\begin{abstract}
In this paper we combine two powerful computational techniques, well-tempered metadynamics and time lagged independent component analysis. The aim is to develop a new tool for studying rare events and exploring complex free energy landscapes. Metadynamics is a well-established and widely used enhanced sampling method whose efficiency depends on an appropriate choice of collective variables. Often the initial choice  is not optimal leading to slow convergence.  However by analyzing the dynamics generated in one such a run with a time-lagged independent component analysis and the techniques recently developed in the area of conformational dynamics, we  obtain much more efficient  collective variables, that are also  better capable  of illuminating the physics of the system. We demonstrate the power of this approach in two paradigmatic  examples. 

\end{abstract}
\maketitle

Molecular Dynamics (MD)  simulations have become pervasive in contemporary science, and are extensively used in fields as diverse as chemistry, biology and material science. Yet in spite of  many successes, the limited time scale that can be explored  in MD simulations severely limits their scope and power.  In many cases the time scale problem results from  the presence of metastable states separated by kinetic bottlenecks that  render  the transitions between such states  so  rare  that  they take place  on a time scale that far exceeds what can be afforded by ordinary  simulation methods.

For this reason, a plethora of enhanced simulation methods have been suggested.  Starting from the pioneering work of Torrie and Valleau\cite{Torrie1977a}, a wide class of such methods  rely on the identification of appropriate collective variables (CVs)\cite{Darve2001,Huber1994,laio2002escaping,wang2001efficient,Barducci:2008aa,Vlssn2016}. The CVs are a set of functions $s(R)$ of the atomic coordinates $R$ that describe those degrees of freedom  whose sampling needs to be accelerated. This latter goal is achieved by adding  to the interaction potential $U(R)$ an appropriately constructed bias  $V(s(R))$, designed so as to accelerate sampling.  Here we shall focus  on one such method, namely well tempered metadynamics (WTMetaD)\cite{Barducci:2008aa,Vlssn2016} that is enjoying increasing popularity and that offers the possibility in one of its variants, usually referred to as  infrequent metadynamics\cite{Tiwary:2013}, to calculate   transition rates from metastable state to metastable state. However  in WTMetaD the rate of convergence  depends on an appropriate choice of CVs especially when it comes to calculating rates.  In an important paper this has been pointed out by Tiwary and Berne\cite{tiwary2016spectral}  who have proposed the spectral gap optimization of order parameters (SGOOP) method, that is based on maximum caliber and that has been shown to improve the quality of an  initial CV guess in a spectacular way.

This  work offers an alternative  to SGOOP based on the signal processing technique Time-lagged Independent Component Analysis (TICA)\cite{molgedey1994separation,perez2013identification,schwantes2013improvements}.  As in the case of Ref.~\citenum{tiwary2016spectral} we start from a CV that is able to push the system  from one metastable state to another albeit sluggishly  and ameliorate it so  as to greatly improve its efficiency. It has been shown in the important work of No\'e and coworkers\cite{perez2013identification} that TICA provides an optimal solution of the variational approach to conformational dynamics (VAC), hence identifying slow order parameters which may serve as optimal CVs in a metadynamics simulation. We shall refer to this combination as the variational conformational dynamics approach to metadynamics (VAC-MetaD). 

We recall first that  in WTMetaD  the bias potential $V(s(R),)$ is built on the fly by periodically adding a small repulsive Gaussian whose amplitude decreases as the simulation progresses.  A remarkable feature of WTMetaD is that this stochastic, apparently out of equilibrium process,  can be described by an ordinary differential equation whose asymptotic solution for the bias tends rigorously to the following equilibrium result\cite{dama2014well}:

\begin{equation}
\label{eq:bias}
\lim_{t \rightarrow \infty } V(s,t)= -\left(1-\frac{1}{\gamma}\right) F(s)
\end{equation}
where $F(s)$ is the free energy associated to the s CV  given within an irrelevant constant by:
\begin{equation}
\label{eq:fes}
F(s)=-\frac{1}{\beta} \log{\int dR \ \delta(s-s(R)) e^{-\beta U(R)}}
\end{equation}
where $\beta = 1/k_BT$ is the inverse temperature and $\gamma$ is the so-called bias factor.

One of the  consequences of the existence  behind WTMetaD of an ordinary differential equation is that the reweighting of the trajectories can be done on the fly and the equilibrium expectation value of an operator $\langle O(R) \rangle $  can be calculated\cite{tiwary2014time}
as an average over the metadynamics run as: 
\begin{equation}
\label{eq:reweigh}
\langle O(R) \rangle = \lim_ {T \rightarrow \infty} 
\frac{\int_0^T O(R_t) e^{\beta(V(s(R_t))-c(t))}dt}{\int_0^T  e^{\beta(V(s(R_t))-c(t))}dt}
\end{equation} 
where $R_t$  are the atomic positions at time $t$ 
and the time dependent energy offset  $c(t)$ is: 
\begin{equation}
\label{eq:c(t)}
c(t)=-\frac {1}{\beta} \log \frac {\int ds e^{-\beta (F(s)+V(s,t))}}{\int ds e^{-\beta F(s)}} 
\end{equation}
a quantity that tends asymptotically to the reversible work performed on the system by the bias. If we  introduce a new time scale $\tilde{t}$ such that
 \begin{equation}
 \label{eq:tau}
 d\tilde{t} = e^{\beta(V(s(R_t))-c(t))}dt 
\end{equation}
we can recognise that  $\langle O(R) \rangle $ can be written as an average over the  $\tilde{t}$  time
\begin{equation}
\label{eq:tauav}
\langle O(R) \rangle = \lim_ {T_{\tilde{t}} \rightarrow \infty} 
\frac{1}{T_{\tilde{t}}}\int_0^{T_{\tilde{t}}} O(R_{\tilde{t}}) d\tilde{t}
\end{equation}
where $T_{\tilde{t}} =\int_0^T  e^{\beta(V(s(R_t))-c(t))}dt$ is the total elapsed $\tilde{t}$ time.
The times $t$ and $\tilde{t}$  measure the metadynamics and  Boltzmann sampling progress respectively.
It follows from Eqn. \ref{eq:tauav} and from the convergency properties of WTMetaD that we can think of WTMetaD as an ergodic dynamics in $\tilde{t}$ time that samples the Boltzmann distribution.
We note however that the $\tilde{t}$ dynamics cannot be directly related to the unbiased dynamics but depends on the choice of CVs. Poor CVs  lead to long  convergence times, while good CVs lead to much shorter ones. In fact the very purpose of biasing the system is to turn rare events into frequent ones, and the time $\tilde{t}$ should not be confused with the actual unbiased time, but a measure of the extent of metadynamics enhancement of the sampling. 

We want now to take advantage of progress made in the field of conformational dynamics\cite{schutte1999direct,prinz2011markov}  and the realization  that a propagator associated to an ergodic dynamics   can be spectrally decomposed into eigenvalues $\lambda_i (t)$ and eigenfunctions $\psi_i(R)$. The highest eigenvalue $\lambda_0$ is one since the system evolves towards the  invariant distribution, while the others decay with time\cite{schutte1999direct}.  If the first sets of M non trivial eigenvalues are separated by a gap from all others, the corresponding eigenfunctions can be identified  as the CVs that describe the slowest modes of the system. In order to practically benefit from these theoretical results one needs an approximation method that can deal with the fact that the propagator is very high dimensional. Luckily  an  approximate evaluation, based on a variational principle similar to the Raleigh-Ritz principle of quantum mechanics, has  been suggested\cite{noe2013variational,nuske2014variational}.  In this approach the dynamics  is projected into a low dimensional space spanned  by the basis functions $O_k (R) \quad k=1, ..., N$ and the eigenfunctions are approximated by a linear expansion:
\begin{equation}
\psi_i(R)=\sum_{k=1}^N b_{ik} O_k(R)
\label{eq:eif}
\end{equation}
The eigenvalues and eigenfunctions that best approximate  the exact eigenvalues and eigenvectors are given by the solution  of the following generalized eigenvalue equation: 
\begin{equation}
{\bf C }(\tau) {\bf b}_i={\bf C }(0)\lambda_i (\tau) {\bf b}_i 
\label{eq:tica}
\end{equation}
where $\tau$ is the lag time and ${\bf C }(\tau)$  is the matrix of the dynamical correlation functions $C_{j,k}(\tau)=\langle O_j(t)O_k(t+\tau) \rangle$,while  $\lambda(\tau)_i$  are the eigenvalues and ${\bf b}_i$ are the expansion coefficients of eigenfunctions.

As discussed above, it is possible to map WTMetaD into  a dynamics that asymptotically samples the Boltzmann distribution.  Thus we  make the ansatz  that also  for  the $\tilde{t}$ dynamics   the properties at the basis of the spectral decomposition described above hold at least asymptotically since metadynamics in this limit explores ergodically the Boltzmann distribution. 

If this is so  we can approximate  the slow modes of the systems with the solutions of  the generalized eigenvalue Eqn. \ref{eq:tica}  in which the time correlation functions are expressed as a function of the scaled time $\tilde{t}$ .  While the variational approach Eqn. \ref{eq:tica} aims at identifying the slowest CVs by varying ${\bf b}_i$ so as to maximize the eigenvalues $\lambda(\tau)$, metadynamics aims at generating a biased dynamics in which the slowest processes are fast. Hence, a successful application of VAC-MetaD should result in a biased dynamics whose slowest processes are fast, corresponding to small leading eigenvalues, $\lambda_1, \lambda_2, ...$.
Thus our strategy will  be   of   choosing first a set of  CVs expressed in the space spanned  by an appropriate set of functions  $O_k(R)$  and then perform a WTMetaD run to calculate the correlation function in $\tilde{t}$ time, plug them into Eqn. \ref{eq:tica}, and use the eigenvectors  of the highest eigenvalues as  improved CVs.  If  we  perform a new  WTMetaD simulation driven by such CVs  then we should see  a much more efficient sampling, and due to the acceleration of the slowest modes, we should observe a decrease in the relaxation time associated with the leading eigenvalues. 
We must add however that since metadynamics needs some incubation time $\tau_c$ before reaching the asymptotic limit in which Eqn. \ref{eq:reweigh} holds, only after time $\tau_c$ will the trajectory yield a stable estimate of the eigenvalues. Only in this limit the eigenvectors of the slowest modes in Eqn. \ref{eq:tica} will be used as new CVs.

We now explicate how the method works in practice. As stated above the VAC-MetaD procedure aims to find from an initial set of candidate order parameters $\textbf{O}=\{O_k(R)\}$ the optimal linear combination of these order parameters to use as a CV in metadynamics. Initially we give equal weight to each order parameter and perform a short biased simulation with the initial CV as $s^{(0)}(R)=\frac{1}{\sqrt{N}}\sum^N_k O_k(R)$.
In order to approximate the unbiased slow modes from the biased metadynamics simulation, we need to reweight trajectory samples in our computation of the correlation matrices\cite{wu2017variational}. 
The two correlation matrices needed to solve Eqn. \ref{eq:tica} are 
\begin{eqnarray}
\textbf{C}(0)&=&\sum_{t}w(t) \textbf{O}(t) \textbf{O}(t)^T \nonumber \\
\textbf{C}(\tau)&=&\sum_{t}w(t)\textbf{O}(t) \textbf{O}(t+\tau)^T
\label{EQ:Corrmatrix}
\end{eqnarray}
with $\textbf{O}(t)$ the set of candidate order parameters at time $t$ and $w(t)$ the WTMetaD weight 
\begin{equation}
w(t)=\frac{e^{\beta(V(s(t))-c(t))}}{\sum_t e^{\beta(V(s(t))-c(t))}}
\end{equation}
In practice the correlation functions of Eqn. \ref{EQ:Corrmatrix} are symmetrized to ensure that the $\lambda$s are real valued\cite{wu2017variational}. Diagonalization of the correlation matrix $\textbf{C}(0)$ would give the usual reweighted principle components.
In the time-lagged matrix $\textbf{C}(\tau)$ the lag time $\tau$ is given by the sum of the rescaled time steps 
\begin{equation}
\tau=\sum_{t=t_0}^{\tau'}e^{\beta(V(s(t))-c(t))} \Delta t
\end{equation}
Insertion of the matrices given by Eqn. \ref{EQ:Corrmatrix} into Eqn. \ref{eq:tica} and solving the generalized eigenvalue equation for the eigenfunctions ${\bf b}_i$ gives a new set of N transformed basis functions $s_i={\bf b}^T_i \textbf{O}$, which are an approximation for the eigenfunctions of the dynamical propagator (Eqn \ref{eq:eif}). 
Each of the N eigenvalues has an associated relaxation time given at a fixed chosen lag time $\tau$ by 
\begin{equation}
t^*_i=-\frac{\tau}{\log{|\lambda_i|}}
\label{EQ:relaxtime}
\end{equation}
One expects to find a gap in timescales between the the $M < N$ slow modes corresponding to large $t^*$ which provides a natural way to select a subset of the N basis function, those corresponding to the slowest relaxation times (largest eigenvalues), as new CVs to be biased with WTMetaD. Thus the procedure yields the desired set of M CVs as
\begin{equation}
s_i(R)=\sum_{k=1}^{M <N} b_{ik} O_{k}(R)
\label{eq:CV}
\end{equation}
with $\textbf{b}_i$ the expansion coefficients in Eqn. \ref{eq:tica}. 
\begin{center} 
\begin{figure}[]
\includegraphics[scale=0.35]{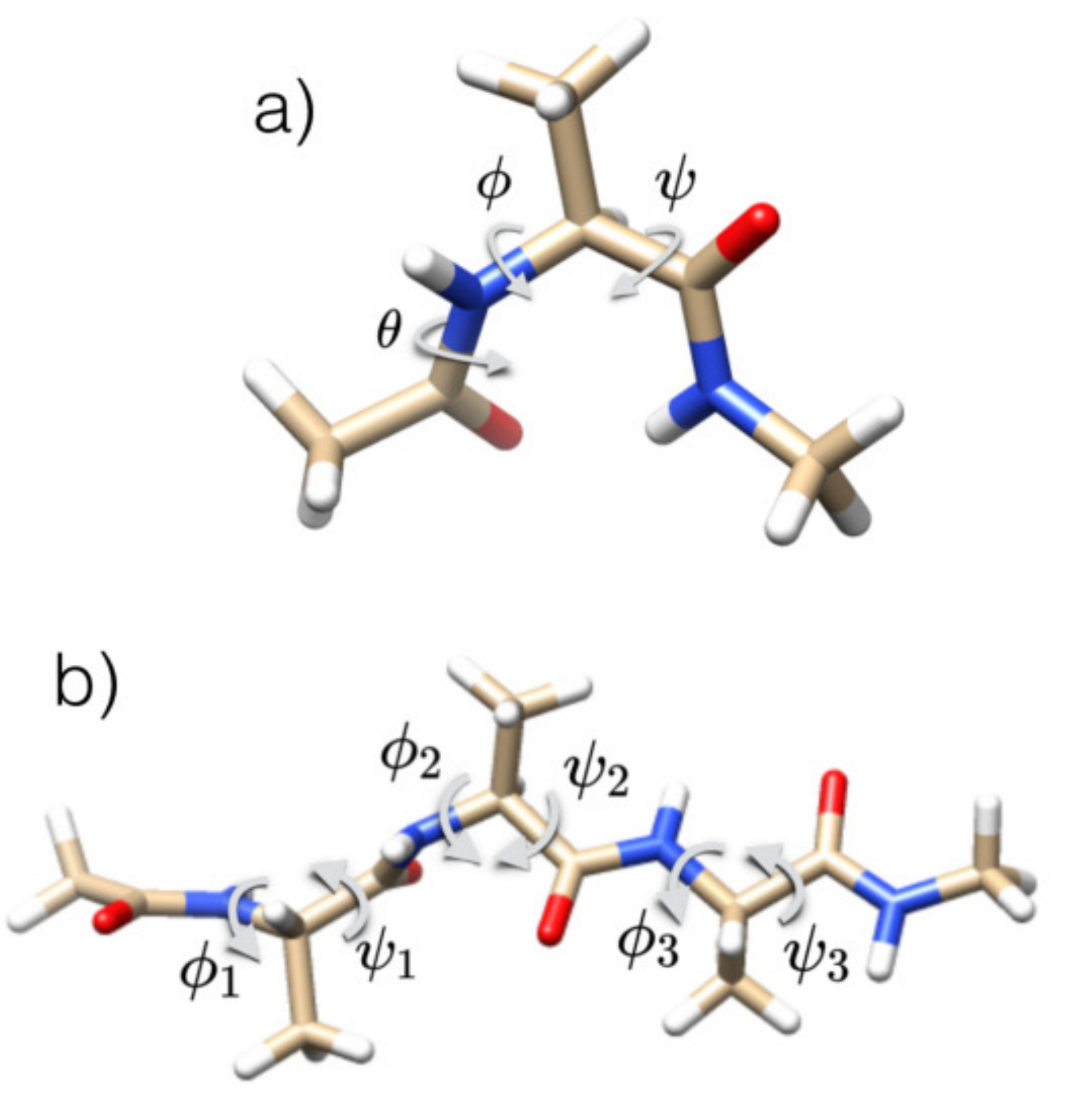}
\caption{Structure and dihedral angle definitions for the two model systems studied: a) alanine dipeptide and b) alanine tetrapeptide}
\label{FG:Structures}
\end{figure}
\end{center}

We shall test this procedure by studying in vacuum the conformational landscape of two simple peptides that have often been used as testing ground for new methods, alanine dipeptide (Ace$-$Ala$-$Nme) and alanine tetrapeptide (Ace$-$Ala$_3$$-$Nme) which are shown in Fig. \ref{FG:Structures}. All simulations were performed with the GROMACS5.0.5 package\cite{hess2008gromacs}  using the Amber99-SB force field with an integration time step of 2 fs. Trajectories were generated in the NVT ensemble with the temperature maintained at 300 K using the stochastic velocity rescaling thermostat\cite{bussi2007canonical}. All metadynamics calculations were performed within the PLUMED2 plugin\cite{tribello2014plumed} with Gaussian hills deposited every 500 integration steps and an initial hill height of 1.2 kJ/mol, a width of 0.03 units and a bias factor $\gamma$ of 15. 

\begin{center}
\begin{figure*}[hbt!]
\includegraphics[width=\textwidth]{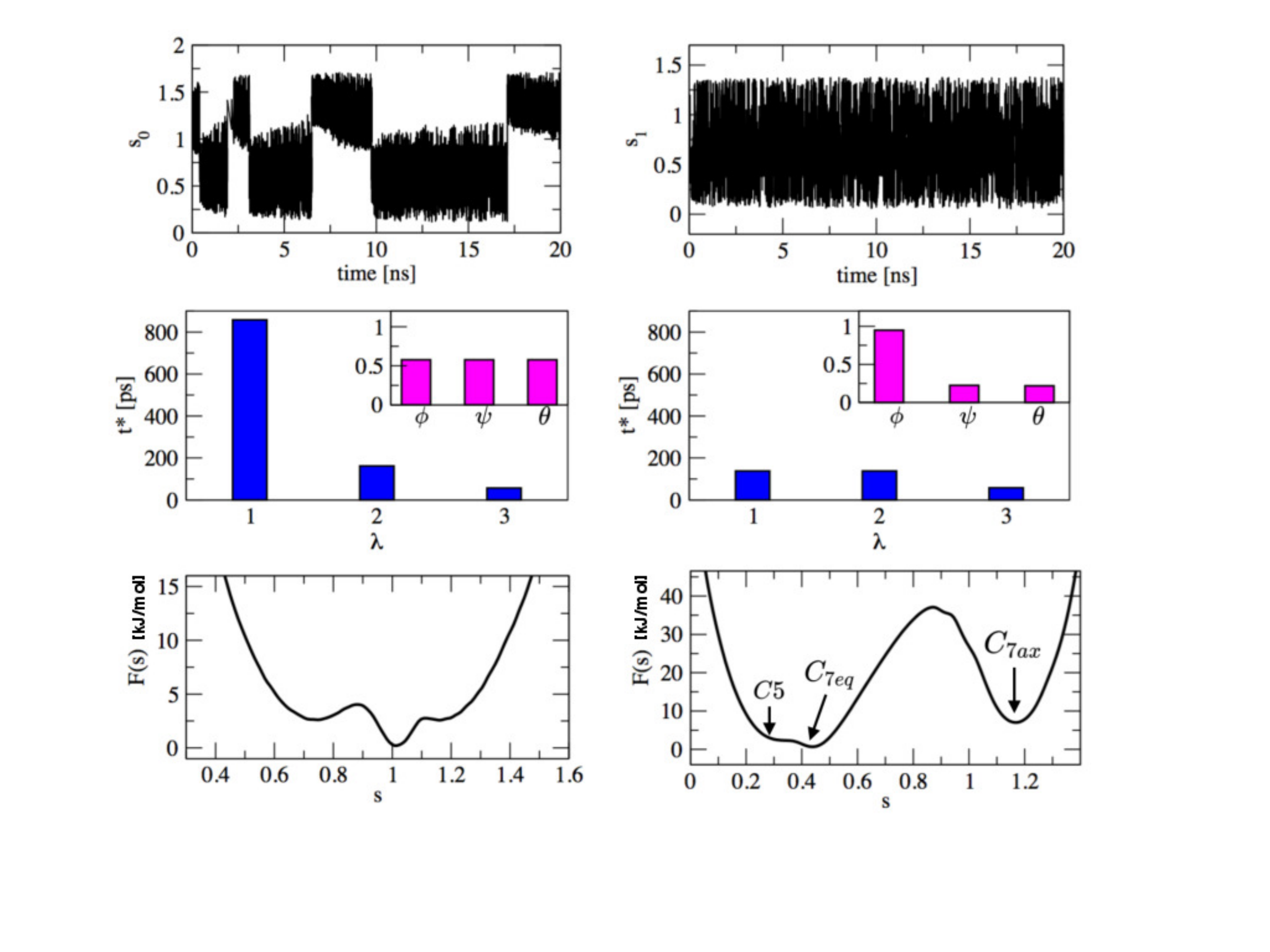}
\caption{Trajectory of alanine dipeptide obtained from a WTMetaD simulation biasing Left Column: initial CV and Right Column: optimized CV from VAC-MetaD. Top Row: Time series of the biased CV for the first 20 ns. Middle Row: The relaxation times of the VAC-MetaD eigenvalues after trajectory reweighting. The initial CV shows a slow relaxation process whereas with the optimized CV all processes are fast. The coefficients corresponding to the initial guess (left) and optimized CV (right) are shown in the inset. Bottom Row: The reweighted free energy surface as a function of the biased CV. The optimized CV (right) clearly distinguishes between the metastable states.}
\label{FG:AlanineDipeptide}
\end{figure*}
\end{center}

We start with the much studied case of alanine dipeptide. Here we depart from the usual procedure and besides the Rahmachandran angles  $\phi$ and $\psi$, we consider additionally the angle $\theta$ that is known to be part of the reaction coordinate\cite{bolhuis2000reaction} (see Fig. \ref{FG:Structures} a). Each angle is transformed according to $\Theta_k=0.5+ 0.5 \cos(\Theta_k - \Theta_0)$ with the reference angle $\Theta_0=1.2$ rad\cite{tiwary2016spectral} and $\boldsymbol{\Theta}=\{\phi,\psi,\theta\}$. The initial CV is then taken to be a linear combination with equal weights of the transformed angles $\phi$, $\psi$, and $\theta$, i.e. $s_0=c_1 \Theta_\phi +c_2 \Theta_\psi + c_3 \Theta_\theta$ with the $\{ c_1,c_2,c_3\}$ initially all equal and normalized to 1. An upper and lower restraint on the $\theta$ angle was introduced at $\pm 0.5$ radians. Following the usual TICA procedure we subtract the reweighted mean given by Eqn. \ref{eq:reweigh} from the raw trajectory data and work in a zero-mean vector space. From an initial WTMetaD trajectory of 20 ns biasing $s_0$ we compute the two correlation matrices in Eqn. \ref{EQ:Corrmatrix} with $\textbf{O}=\boldsymbol{\Theta}$.
The trajectory of the initial CV $s_0$ is shown in Fig. \ref{FG:AlanineDipeptide} (top left). It can be seen that such a CV is able to induce transitions between the different conformers however such transitions are not very frequent.  This is reflected by the fact that the highest $\lambda (\tau)$ decays more slowly than the other eigenvalues reflecting a difficulty of the chosen CV to promote transitions. The middle row of Fig. \ref{FG:AlanineDipeptide} shows the relaxation times $t^*_i$ associated with each eigenvalue given by Eqn. \ref{EQ:relaxtime} at a lag time $\tau=800$ ps chosen within the regime for which the eigenvector coefficients are stable. Fig. \ref{FG:AlanineDipeptide} middle left clearly shows a slow process with a dominant large eigenvalue and a clear separation of time scales. The eigenvector associated with this highest eigenvalue, computed in the asymptotic regime where the eigenvectors are constant with respect to the lag time $\tau$, thus obtained is used as a new CV $s_1 = \textbf{b}_1^T \boldsymbol{\Theta}$. If we use as CV this eigenvector associated with the highest $\lambda$  the exploration of the conformational space is greatly accelerated as depicted in Fig. \ref{FG:AlanineDipeptide} top right, and  the relaxation rates of decay of the different $\lambda$s become  comparable, reflecting the fact  that the rare event has been made no longer rare (Fig. \ref{FG:AlanineDipeptide} middle right). It is interesting to note that in the optimized CV much of the weight is of the angle $\phi$ but both $\psi$ and $\theta$ are part of the optimal CV.

In Fig. \ref{FG:AlanineDipeptide} (bottom) we also show the free energies associated to the initial (left) and final (right) CV. The conformational landscape of alanine dipeptide is well known and characterized by a deeper basin in which conformers C5 and C7eq can easily interconvert and while conformer C7ax is higher in energy separated by a sizeable barrier from the first two. It can be seen that although the FES is fully converged, these physical pictures are hidden in the free energy representation provided by the initial CV but clearly evident when using the final CV.

\begin{center}
\begin{figure*}[hbt!]
\includegraphics[scale=0.5]{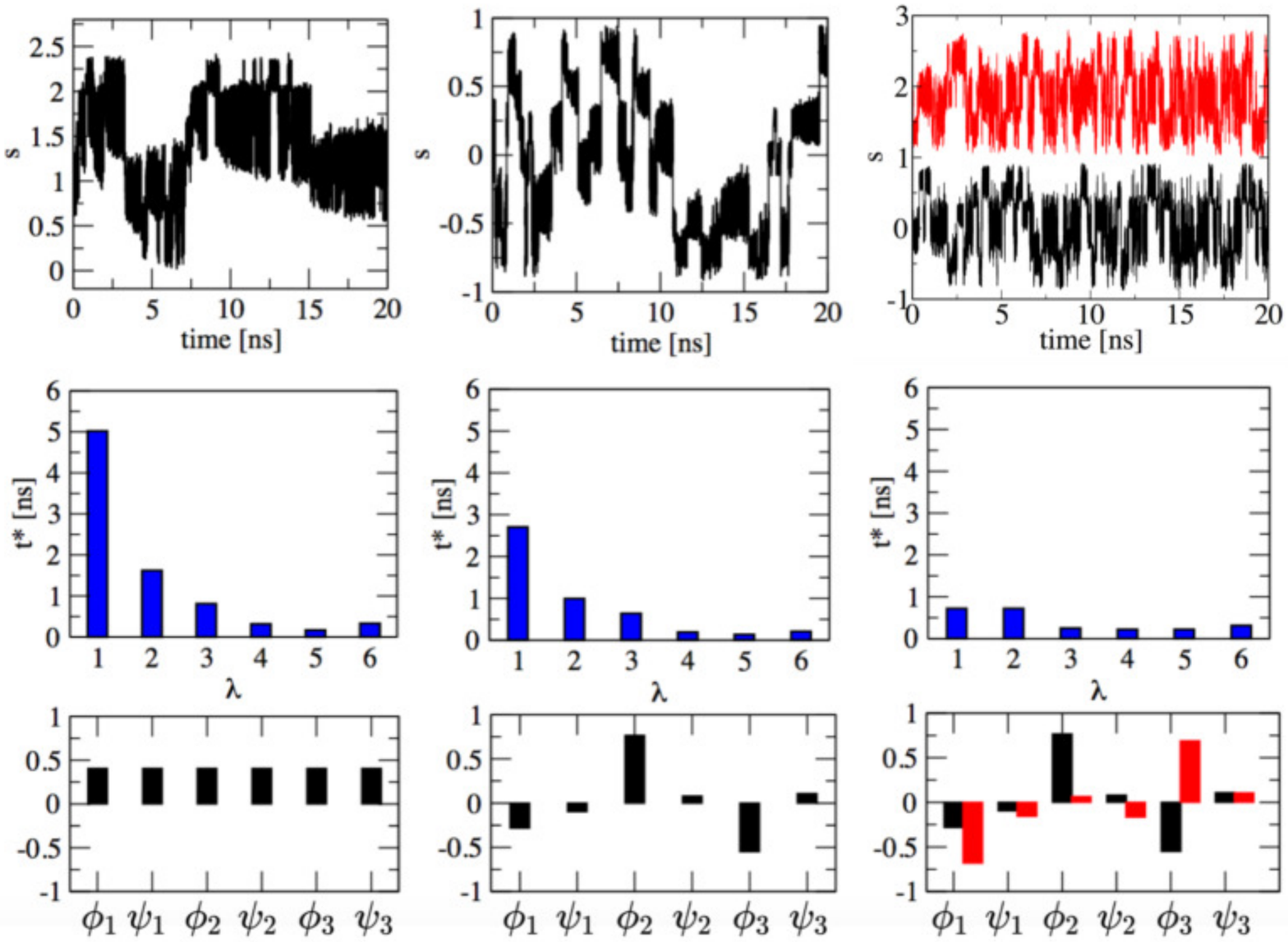}
\caption{Trajectory of alanine tetrapeptide obtained from a WTMetaD simulation biasing Left Column: initial CV, Middle Column: first VAC-MetaD eigenvector only, and Right Column: both the  first and second VAC-MetaD eigenvector used as CV. [The second CV shown in red is shifted for clarity] Top Row: Time series of the biased CV for the first 20 ns. Middle Row: Relaxation times of the VAC-MetaD eigenvalues after trajectory reweighting. [relaxation times and eigenvectors are shown for $\tau=1$ ns for which the eigenvectors are stable.] Bottom Row: The coefficients of each angle used to define the biased CV. The middle and right coefficients are obtained from the solution to Eqn \ref{eq:tica} from the initial trajectory. In the far right column the first CV $s_1$ is shown in black and the second CV $s_2$ is shown in red.}
\label{FG:Ala3-A}
\end{figure*}
\end{center}

\begin{figure*}[hbt!]
\includegraphics[width=\textwidth]{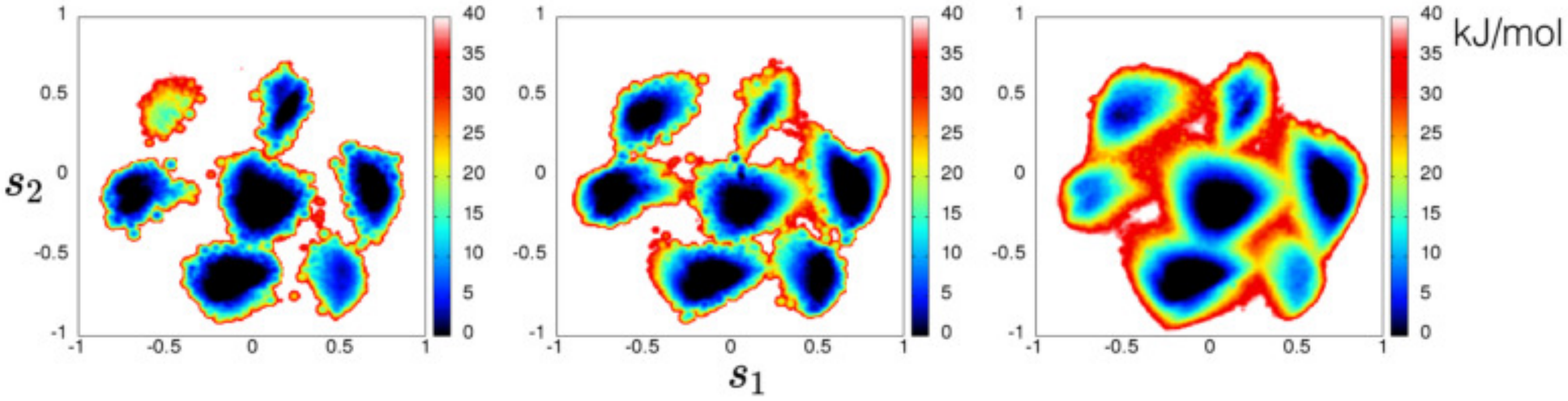}
\caption{Reweighted free energy surface as a function of the first two VAC-MetaD eigenvectors for alanine tetrapeptide from a trajectory biasing the Left: initial guess CV, Middle: first eigenvector only, and Right: first and second eigenvectors.}
\label{FG:Ala3-B}
\end{figure*}

We now turn to the second example for which we have applied our approach, that of alanine tetrapeptide (Ala$_3$) shown in Fig. \ref{FG:Structures} b), a case already considered in\cite{valsson2014variational,tiwary2016spectral}. The simulation is started as in Ref. ~\citenum{tiwary2016spectral} by taking as collective coordinate a linear combination with equal weights of the set of six dihedral angles $\boldsymbol{\Theta} =\{ \phi_1,\psi_1, \phi_2, \psi_2, \phi_3, \psi_3 \}$ transformed as before so that $\Theta_k=0.5+ 0.5 \cos(\Theta_k - \Theta_0)$.  As can be seen in Fig. \ref{FG:Ala3-A} (left) the exploration of the conformational space is somewhat inefficient with rare conformational transitions. We project the $\tilde{t}$ dynamics in the space of the chosen angles and observe that the decay times of the two topmost eigenvalues are significantly slower than that of the others. This suggests to project the free energy surface in the space of the two topmost eigenvectors. As shown in Fig. \ref{FG:Ala3-B}, in this representation the seven different minima identify conformers that have a  different arrangement of the three dihedral angles $\phi_1$, $\phi_2$, and $\phi_3$. The  conformer  in which all the three $\phi$s have all  positive values  is seldom visited and in this representation are projected into the top left tail of the central minimum.  
 
An improvement in sampling efficiency is obtained if one uses the topmost eigenvector as a new CV (see the central panels in Fig. \ref{FG:Ala3-A}), but still it can be seen that a relatively slow process remains.  However the behavior in time of the $\lambda$s cries out for the use of at least two CVs. If this is done, (see Fig. \ref{FG:Ala3-A} far right panels) the improvement is amazing and it offsets the cost of using two CVs instead of one. In fact within a 20 ns  run we get better converged results than the  extensive parallel tempering run in Ref ~\citenum{valsson2014variational} that used an aggregated time equivalent to $8 \times 500$ ns. These lead to a well converged and smooth free energy surface (see Fig. \ref{FG:Ala3-B} right).

It is difficult at this stage to compare the relative practical merits of the SGOOP method to ours. Extensive test on a variety of applications will be needed. As far as we can tell, for the two cases examined here they seem to have comparable performances in the case of Ala$_3$ when we use only the topmost eigenvector as collective variable. It is likely that the two methods will be complementary. However a difference in philosophy must be underlined. In SGOOP one reweighs a one-dimensional projection of the FES to find the optimal linear combinations of CVs. Here we are reweighting the simulation time to take advantage of the variational formalism of conformation dynamics whose solution provides an optimal estimate for the true dynamical propagator. 

In summary, there is a growing interest in applying dimensionality reduction techniques on a larger candidate set of possible CVs to find a subset of generic good CVs that can be used for enhanced sampling. A widely used example is principle component analysis (PCA) which projects the data along the direction of largest variance. On the other hand, it is well known in the field of conformational dynamics that time-lagged independent component analysis on high dimensional coordinates from molecular dynamics can be useful for constructing Markov state models. In this letter we have taken this insight and combined it with well tempered metadynamics to find a set of optimal CVs for further enhanced sampling. We surmise finally that the method can be adapted to other sampling methods  by appropriately changing the weights in Eqn. \ref{EQ:Corrmatrix}

\begin{acknowledgments}
While preparing this manuscript a work has appeared in the literature\cite{m2017tica} in which it is pointed out that from the TICA formalism useful collective coordinates can be extracted, once a long enough trajectory is available from unbiased simulations. Our work extends considerably the scope of their approach by making it applicable to the more commonly encountered situation in which transitions between metastable states can only be observed by the use of a biased simulation.
The authors thank Frank No\'e and Pratyush Tiwary for careful reading of the manuscript and useful suggestions. Computational resources were provided by the Swiss National Supercomputing Center (CSCS). This research was supported by the VARMET European Union Grant ERC-2014-ADG-670227 and the National Center of Competence in Research Materials Revolution:
Computational Design and Discovery of Novel Materials (MARVEL)
51NF40\_141828.
\end{acknowledgments}

\bibliography{VACMeta}

\end{document}